\documentclass[a4paper,fleqn,usenatbib]{mnras}

\usepackage{newtxtext,newtxmath}

\usepackage[T1]{fontenc}
\usepackage{ae,aecompl}

\usepackage{graphicx}   
\usepackage{amsmath}    
\usepackage{amssymb}    
\usepackage{makecell}   
\usepackage{threeparttable} 

\title[Correlation between the continuum and \ion{C}{iv} BALs]{Correlation between the ionizing continuum and variable \ion{C}{iv} broad absorption line in multi-epoch observations of SDSS J141007.74+541203.3}

\author[H.-Y. Huang et al.]{
Hong-Yan Huang,$^{1}$\footnotemark[1]
Cai-Juan Pan,$^{2}$\footnotemark[1]
Wei-Jian Lu,$^{3}$
Yi-Ping Qin,$^{4}$
Ying-Ru Lin,$^{3}$
\newauthor
{ }Wei-Rong Huang,$^{4}$
Yu-Tao Zhou,$^{5,6}$
Min Yao,$^{3}$
Wei-Jing Nong,$^{2}$
Mei-Mei Lu,$^{2}$
\newauthor
{ }Zhi-Kao Yao$^{2}$
and Qing-Lin Han$^{2}$
\\
$^{1}$School of Physics and Electronic Information, Yunnan normal University, Kunming 650500, China\\
$^{2}$School of Materials Science and Engineering, Baise University, Baise 533000, China\\
$^{3}$School of Information Engineering, Baise University, Baise 533000, China\\
$^{4}$Center for Astrophysics, Guangzhou University, Guangzhou 510006, China\\
$^{5}$Key Laboratory of Optical Astronomy, National Astronomical Observatories, Chinese Academy of Sciences, Beijing 100012, China\\
$^{6}$University of Chinese Academy of Sciences, Beijing 100049, China}

\date{Accepted 2019 XXX. Received 2018 YYY; in original form 2018 ZZZ}

\pubyear{2017}

\begin{document}
\label{secondpage}
\pagerange{\pageref{secondpage}--\pageref{lastpage}}
\maketitle

\begin{abstract}
Correlation between the variations of quasar absorption lines and the ionizing continuum have been recently confirmed in systematic studies. However, no convincing individual case is reported. We present a statistical analysis of the variable \ion{C}{iv} broad absorption line (BAL) in the quasar SDSS J141007.74+541203.3, which have been observed with 44 epochs by the Sloan Digital Sky Survey Data Release 14. \citet{Grier2015} has recently concluded that the most likely cause of the variability of the BAL in SDSS J141007.74+541203.3 is a rapid response to changes in the incident ionizing continuum. In this paper, we confirm the anticorrelation between the equivalent width of BALs and the flux of the continuum based on the spectra of this quasar that show significant variations, which serve as another independent evidence for the view of \citet{Grier2015}.

\end{abstract}

\footnotetext[1]{E-mail: 936499587@qq.com (H-Y, H); pancj2017@126.com (C-J, P)}

\begin{figure*}
\includegraphics[width=2\columnwidth]{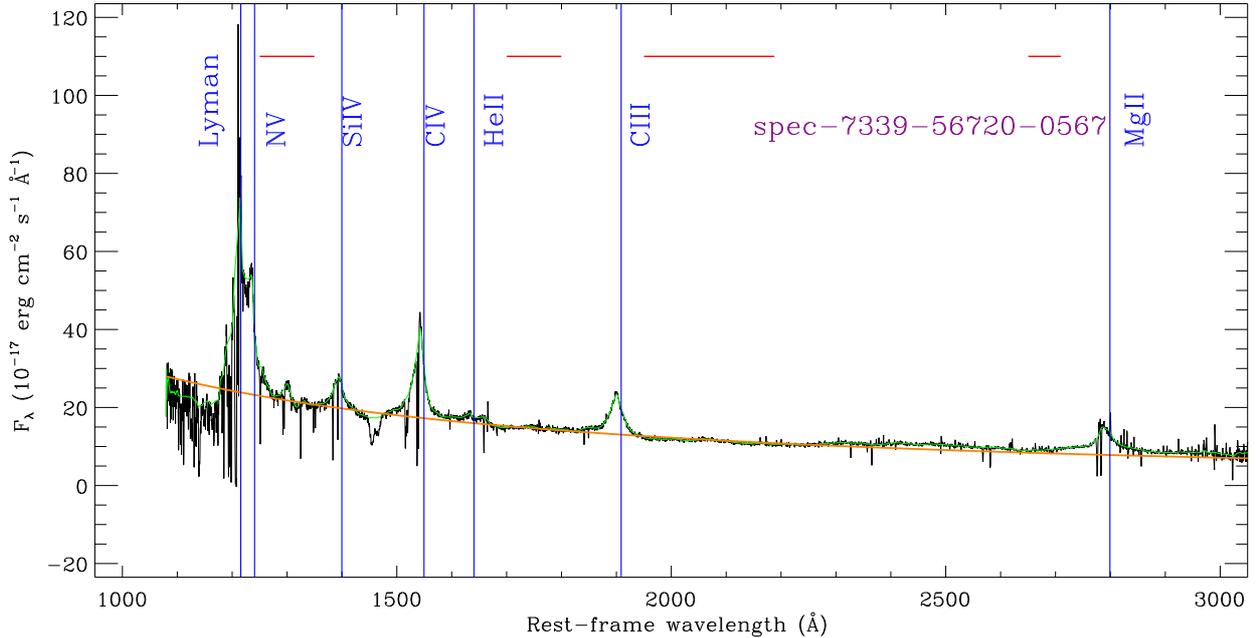}
\caption{The continuum fitting of SDSS J1410+5412 (on MJD 56720). The blue verticals mark the emission lines. The black curve is the original spectrum from the SDSS. The orange curve is the power-law continuum fitting. The green curve is the pseudo-continuum spectrum. The red horizontal lines are the relatively line-free wavelength regions (1250--1350, 1700--1800, 1950--2200, and 2650--2710\,\AA~in the rest frame) for continuous iterative fitting. }
    \label{fig:1}
\end{figure*}

\begin{figure}
\includegraphics[width=\columnwidth]{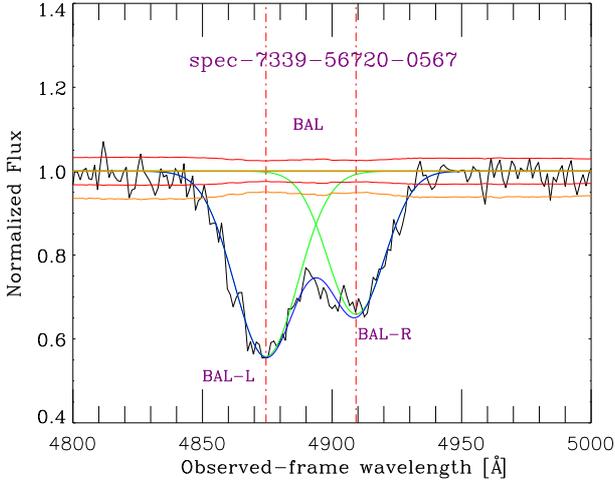}
\caption{The fitting for the \ion{C}{iv} BAL in SDSS J1410+5412 (on MJD 56720). The black curve is the original spectrum from the SDSS. The orange thick line is pseudo-continuum-normalized spectrum; the red and orange curves are the uncertainty levels that have been normalized by the corresponding pseudo-continua; the red vertical dotted lines are the center of the BAL-L and BAL-R trough, respectively; the green curves are the Gaussian fittings for the BAL-L and BAL-R trough; the blue curves are multiple Gaussian fittings for them.}
    \label{fig:2}
\end{figure}

\begin{figure}
    \includegraphics[width=\columnwidth]{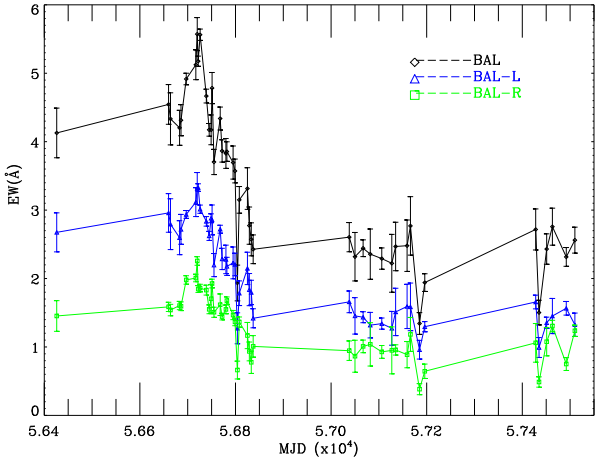}
  \caption{The EW of the BALs versus the time within MJD=56426$\sim$57510 (in the rest frame). The black, blue and green curves are the EWs of BAL, BAL-L, and BAL-R, respectively. }
    \label{fig:3}
\end{figure}
\begin{figure}
	\includegraphics[width=\columnwidth]{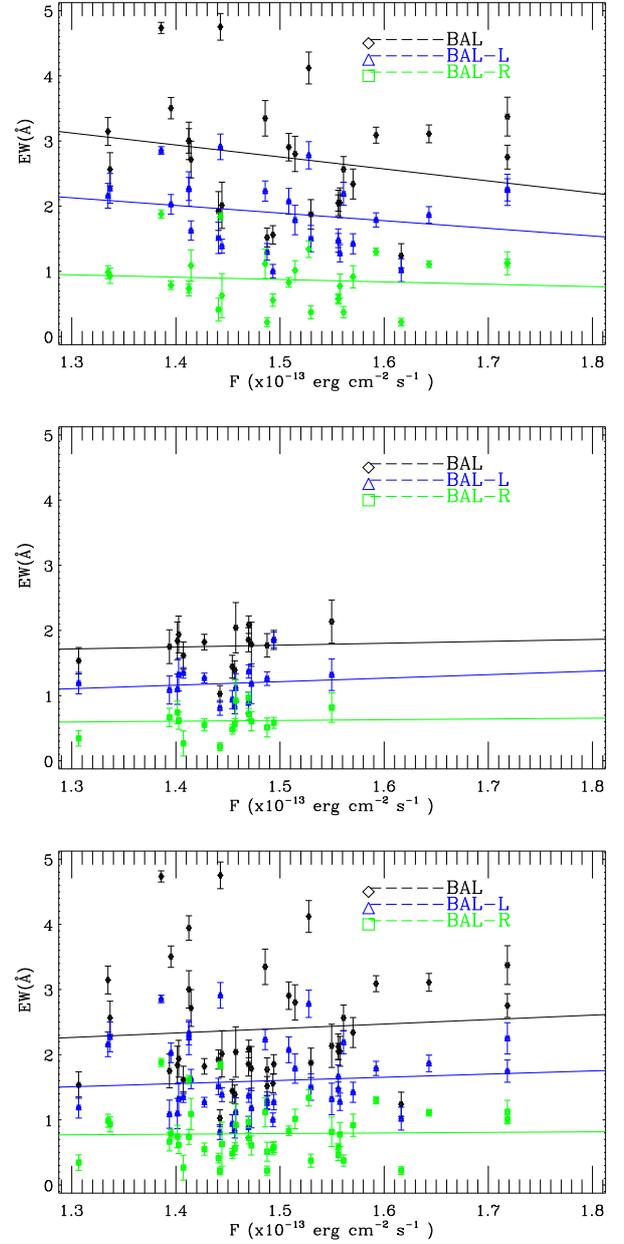}
    \caption{Plots of the EW of BALs versus the flux of the continuum for three samples. In each panel, the black, blue and green plots represent the BAL, BAL-L, and BAL-R, respectively. The solid lines are the corresponding linear regression fits. The top panel represents the variation sample (the MJD from 56426 to 56837). The middle panel represents the no variation sample (the MJDs are from 57038 to 57510). The bottom panel represents the whole sample, which includes all data of 44 epochs (the MJD from 56426 to 57510). For all the whole BALs, BAL-Ls, and BAL-Rs, the moderate anticorrelations between the EW of BALs and the flux of the continuum are found for the variation sample, while no correlation between them for both the no variation sample and the total sample.}
    \label{fig:4}
\end{figure}
\begin{figure}
	\includegraphics[width=\columnwidth]{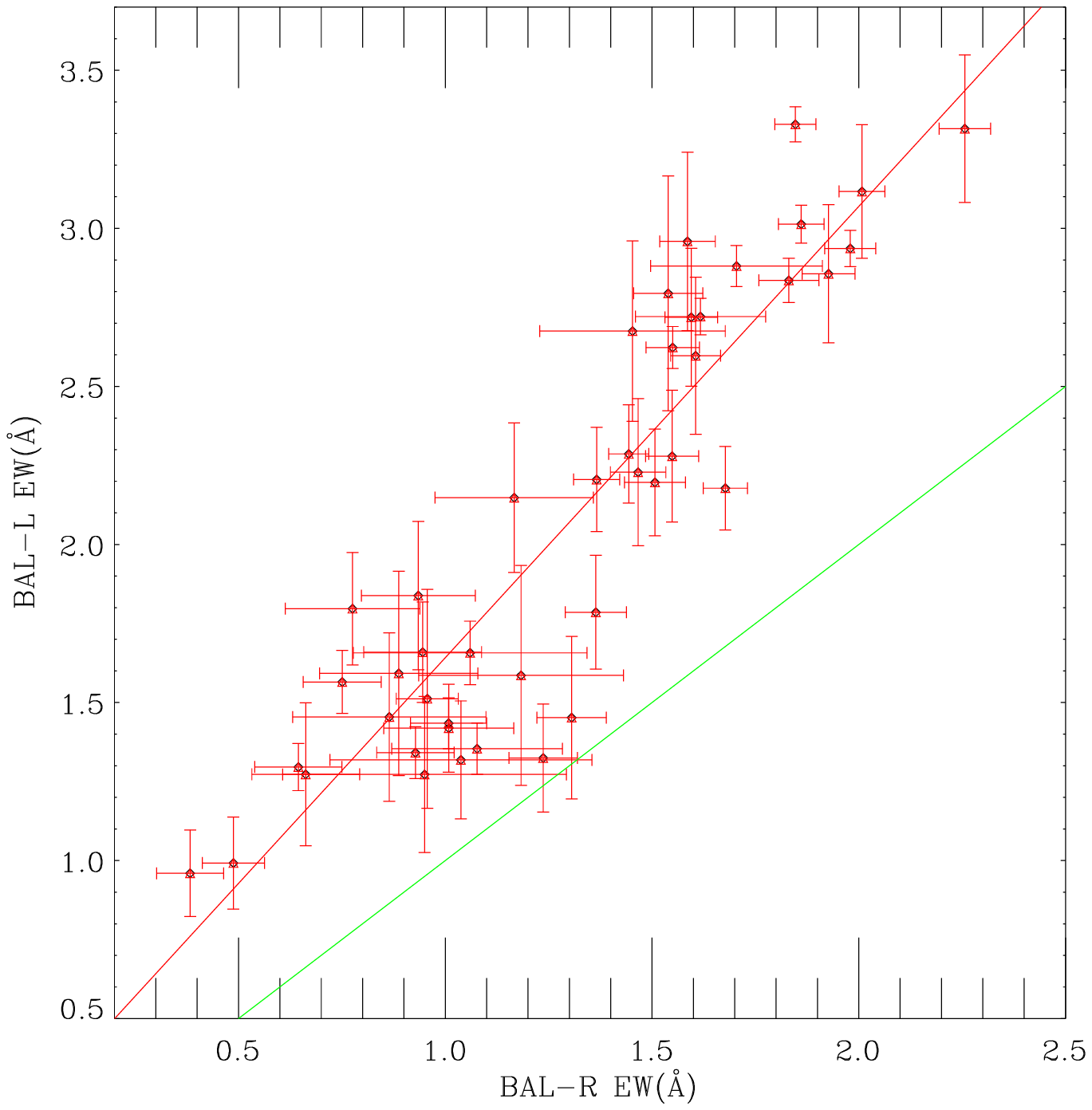}
    \caption{Comparison of the EWs between BAL-L and BAL-R. The red line is the linear regression fit for the data points. The green line is equal EWs for the two absorption troughs.}
    \label{fig:5}
\end{figure}

\section{Introduction}
Quasars are the brightest and densest objects in the active galactic nucleus (AGNs). It is believed that the intrinsic absorption lines are generated by the high-velocity flowing wind and emitted near the supermassive black hole (SMBH) of the quasar. With respect to the full width at half maximum (FWHM, which also be known as width of absorption lines) of line spread, the intrinsic absorption lines are classified as three types: broad absorption lines (BALs) with FWHM$\geqslant$2000 $\rm km~s^{-1}$ (e.g. \citealp{Weymann1991,Hamann2008}); mini-BALs with 2000 $>$FWHM$>$ 500 $\rm km~s^{-1}$ (e.g. \citealp{Misawa2007b,Ganguly2008}); narrow absorption lines (NALs) with FWHM$\leqslant$500 $\rm km~s^{-1}$ (e.g. \citealp{Misawa2007a,Misawa2014}). 

The outflowing winds revealed by intrinsic absorption lines in quasar spectra are important at least for the following two reasons. First, the BALs can be observed in $\sim$ 41 per cent of optically observed quasars (e.g. \citealp{Allen2011}), this observed frequency indicates that outflowing winds play a significant role in the nuclear environment (e.g.\citealp{Crenshaw2003}). Second, as a form of feedback from the quasars into massive galaxies, quasar outflowing winds could remove cold gas and thus regulate star formation and more SMBH accretion (e.g. \citealp{Springel2005,King2010}).

The variation of BALs is a powerful tool for studying the properties of the outflowing gas, such as the size, kinematics, material composition and evolution. The reason for the variations of BAL with time is generally considered to be the transverse motion of the gas (e.g. \citealp{Hamann2008,Shi2016,Hemler2019}) or the ionization variation of the  {outflowing gas} (e.g. \citealp{Grier2015,Lu2017,Lu2018broad,Hemler2019}). Whichever of the above two reasons, BAL variation on short time-scales can be used to measure the distance of the absorber from the central SMBH. Shorter variation time-scales indicate smaller distances. In the traverse scenario, shorter variation time-scales indicate shorter crossing times of the outflowing clouds (\citealp{Hamann2008,Capellupo2011}); while in the ionization change scenario, shorter variation time-scales represent shorter recombination times, which indicates the higher densities of the clouds (\citealp{Hamann1997}). In addition, whether there is variability on longer (several years) time-scales indicates whether the outflowing gas has a persistent structure. Furthermore, variability studies of BAL, mini-BAL and intrinsic NAL can address the evolution of outflows. In an evolution model (e.g. \citealp{Farrah2007}), these three types of absorption lines represent different evolution phases of the outflows. The BAL may be the powerful phase of the outflowing wind, while the other two types of absorption lines may appear at its beginning or the ending stages (e.g. \citealp{Hamann2008}).

Correlation studies between the continuum and the variations of absorption lines have been proposed to investigate the variability reason of absorption lines. Actually, some researches on the correlation between the variations of BALs and the persistence/radiation have been carried out. A few works have reported that no apparent correlations between the variations of BALs and that of the continuum (e.g. \citealp{Gibson2008,Wildy2014,Vivek2014}), but several recent studies have provided evidence of anticorrelations between the variations of absorption lines and the continuum (e.g. \citealp{Lu2018broad,Lu2018saturation} for BAL; \citealp{Lu2017,Chen2018b,Chen2018a} for NAL).

The above-mentioned works are based on the analysis of multiple sources. A particular quasar SDSS J141007.74+541203.3 ($z_{\rm em}=2.35$; \citet{Hewett2010}, hereafter J1410+5412) with multi-epoch observations was explored in detail by \citet{Grier2015} and \citet{Hemler2019}. A linear speed with 4340 $\rm km~s^{-1}$ in rapidly variations of it's \ion{C}{iv} BALs was reported, which is seem to be caused by a rapid response to changes in the incident ionizing continuum. It is interesting to find out whether there is a correlation between the variations of \ion{C}{iv} BAL and the flux of the continuum in J1410+5412. The paper is structured as follows. In Section 2, we describe characteristics and analysis of the data. The correlation analysis and discussion will be given in Section 3, where also including a brief summary. The cosmological parameters used in this article are $\Omega_{\rm M}=0.3$, $\Omega_{\Lambda}=0.7, H_0=70\,\rm km\,s^{-1}\,Mpc^{-1}$.

\section{SPECTRAL ANALYSIS}
The quasar J1410+5412 studied in this paper is a target of the Sloan Digital Sky Survey Reverberation Mapping Project (SDSS-RM; \citealp{Shen2015}),  {the latter is a dedicated multi-target reverberation mapping campaign performed as part of SDSS-III \citep{Eisenstein2011} BOSS survey \citep{Dawson2013}. A technical overview of the SDSS-RM project could be seen in \citet{Shen2015}. The apparent $i$-band magnitude of the quasar is $m_i = 18.1$ \citep{Alam2015}. The average value of the Signal to Noise Ratio (S/N) of the spectra is 26.60 (22.23$\sim$37.54), there are listed in Table~\ref{tab:1}. The spectrograph has wavelengths from 3650 to 10400\,\AA~ and a spectral resolution of R$\sim$2000 \citep{Smee2013}.} We collect 44-epoch spectra of SDSS J1410+5412 from the Data Release 14 (DR14, \citealp{Abolfathi2018}). The time of these observations cross from  MJD 56426 to 57510.

To evaluate the variation of the continuum, we fitted each spectrum of SDSS J1410+5412 using a power-law function.  {The power-law continuum was iteratively fitted  in several wavelength regions (1250--1350, 1700--1800, 1950--2200, 2650--2710\,\AA~in the rest frame), which are defined by \citet{Gibson2009a} as ``relatively line-free (RLF) regions". During the fitting, we ignored the pixels beyond 3$\sigma$ significance for reducing the influences of emission/absorption lines as well as remaining sky pixels (e.g. \citealp{Lu2018complex2}).} An example of the power-law continuum fit is shown in Fig.~\ref{fig:1}. We choose the wavelength range of 1200--2100\,\AA~to obtain the flux of the continuum (F) and the corresponding uncertainty, the formulas are as follows:

\begin{equation}
    F=\displaystyle{\sum_{i}}(\lambda_{i+1}-\lambda_i)\frac{1}{2}(F_{c_{i+1}}+F_{c_i}),
	\label{eq:equa_1}
\end{equation}

\begin{equation}
    \sigma_F=\sqrt{\displaystyle{\sum_{i}} [\sigma_{F_{c_i}}(\lambda_{i+1}-\lambda_{i})]^2}.
	\label{eq:equa_2}
\end{equation}
where $F_c$ is the flux of the power-law continuum, $\sigma_{F_c}$ is the uncertainty of the flux of the power-law continuum, $\lambda$  is wavelength. The values of $F_c$ and $\sigma_{F_c}$ are listed in Table~\ref{tab:1}.

To fit the absorption lines, it is necessary to fit the {pseudo-continuum spectra first.} \textbf{Pseudo-continuum is a fitting curve along the initial spectrum but ignores the absorption lines}. We fit the continua of SDSS J1410+5412 by iteratively performing the cubic spline functions (e.g. \citealp{Lu2018complex1}). To decrease the effects of absorption troughs as well as remaining sky pixels, we masked out the pixels which outside 3$\sigma$ significance from the present fit. The continuum fitting is shown in Fig.~\ref{fig:1}. After that, we measured absorption lines in the spectra that have been normalized by the pseudo-continua. 

As seen in Fig.~\ref{fig:1}, the whole region of \ion{C}{iv} BAL is the combination of two parts (BAL-L and BAL-R). We thus employed two Gaussian functions to fit them, respectively. An example of the Gaussian fitting is displayed in Fig.~\ref{fig:2}. We measured the equivalent widths (EWs) of the absorption lines based on the Gaussian profile, and estimated the error of each Gaussian profile according to the corresponding flux uncertainties via
\begin{equation}
    \sigma_{EW}=\frac{\sqrt{\Sigma_iP^2(\lambda_i-\lambda_0)\sigma^2_{f_i}}}{(1+z_{\rm abs})\times\Sigma_iP^2(\lambda_i-\lambda_0)}{\Delta\lambda}.
	\label{eq:equa_3}
\end{equation}
where P is the Gaussian function at center line $\lambda_0$; $\lambda_i$ is the wavelength of data point; $\Delta\lambda$ is the spacing of adjacent points; $\sigma_{f_i}$ is the uncertainty of the spectral flux of the normalization \citep{Schneider1993}. Then the total EW of the BAL are the sum of the two Gaussian profiles
\begin{equation}
EW_{BAL}=EW_{BAL-L}+EW_{BAL-R}, 
	\label{eq:equa_4}
\end{equation}
and the error of the total EW is
\begin{equation}
  \sigma_{{EW}_{BAL}}=\sqrt{\sigma_l^2+\sigma_r^2}.
	\label{eq:equa_5}
\end{equation}
where $\rm EW_{BAL}$, $\rm EW_{BAL-L}$ and $\rm EW_{BAL-R}$ are the EWs of BAL, BAL-L and BAL-R, respectively; $\sigma_{{EW}_{BAL}}$, $\sigma_l$ and $\sigma_r$ are the errors of BAL, BAL-L and BAL-R, respectively.

\section{correlation analysis AND DISCUSSION}
Fig.~\ref{fig:3} shows the EW measurements of BAL as a function of MJD. It is obviously that the BAL in J1410+5412 shows significant variation during MJD 56426 to 56837, while shows no significant variation during MJD 57038 to 57510. According to this phenomenon, we divided the multi-epoch spectra of J1410+5412 into two samples, i.e. the variation sample (MJD from 56426 to 56837) and the no variation sample (MJD from 57038 to 57510), and performed Spearman correlation to this two samples separately. We also performed a Spearman correlation to the total spectra sample. The statistical parameters are listed in Table~\ref{tab:2}. We plotted the EW of BALs versus the flux of the continuum in Fig.~\ref{fig:4}. The moderate anticorrelation between the EW of BALs and the flux of the continuum was confirmed by the Spearman rank correlation analysis for the variation sample, while no correlation between them for both the no variation sample and the total sample. When we divided each of the BALs into two parts, i.e. BAL-L and BAL-R, as we did in section 2, the correlation tests for the three sample also returned the same results with the total BALs (i.e. moderate anticorrelation between the EW of BAL-Ls/BAL-Rs and the flux of the continuum for the variation sample, while no correlation between them for both no variation sample and the total sample, see also Table~\ref{tab:2} for details) .

These results provide a new clue to understand the variation mechanism of the BAL in J1410+5412. Although there is substantial scatter in the plots, in the variation sample we find an anticorrelation between the the EW of BALs and the flux of the continuum, which serve as a direct evidence for the idea that BAL variability is driven mainly by changes in the gas ionization in response to the continuum variations. The moderate anticorrelation between the EW of BALs and the flux of the continuum of the variation spectra sample can roughly indicate the ionization states of absorbers, because photoionization simulations have shown that when the ionization parameter (U) increasing, the EWs of \ion{C}{IV} rise till a peak, and decrease after then (e.g. \citealp{Wang2015,He2017}). Thus, the absorbers in J1410+5412 are at a relatively high ionization state.

\citet{Grier2015} also hold the same view with us. They drew their conclusions mainly based on the following two points. On the one hand, they found that variations occur across the whole trough rather than in some segments. On the other hand, the high-velocity \ion{C}{IV} BAL (Trough A in their paper) and the mini-BAL (Trough B in their paper) vary coordinately. We plotted the EW of BAL-L versus BAL-R in Fig.~\ref{fig:5}. The strong correlation between them was confirmed by the Spearman rank correlation analysis (see also Table~\ref{tab:2}). This result also confirms the coordinated variability between different parts of BAL trough in J1410+5412.

In summary, we found the correlation between the EW of BALs and the flux of the continuum of the variation spectra sample (MJD from 56426 to 56837), which confirms that the variation mechanism of the BAL in J1410+5412 is the ionization changes of the absorbers, as a response to the continuum variations. Our results support the conclusions made by \citet{Grier2015}.

\section*{Acknowledgements}
We are grateful to the editor and reviewer of this journal for their useful comments. Our work was supported by the Guangxi Natural Science Foundation (No. 2017GXNSFAA198348). QYP acknowledges support from the Excellent Youth Foundation of Guangdong Province (Grant No. YQ2015128), and the Guangzhou Education Bureau (Grant No. 1201410593). 

Funding for SDSS-III was provided by the Alfred P. Sloan Foundation, the
Participating Institutions, the National Science Foundation and the US
Department of Energy Office of Science. The SDSS-III website is
\url{http://www.sdss3.org/.}

SDSS-III is managed by the Astrophysical Research Consortium for the
Participating Institutions of the SDSS-III Collaboration, including the
University of Arizona, the Brazilian Participation Group, Brookhaven National Laboratory, Carnegie Mellon University, University of Florida, the French Participation Group, the German Participation Group, Harvard University, the Instituto de Astrofisica de Canarias, the Michigan State/Notre Dame/JINA Participation Group, Johns Hopkins University, Lawrence Berkeley National Laboratory, Max Planck Institute for Astrophysics, Max Planck Institute for Extraterrestrial Physics, New Mexico State University, New York University, Ohio State University, Pennsylvania State University, University of Portsmouth, Princeton University, the Spanish Participation Group, University of Tokyo, University of Utah, Vanderbilt University, University of Virginia, University of Washington, and Yale University.





\bibliographystyle{mnras}
\bibliography{hy}

\begin{thebibliography}{}
\makeatletter
\relax
\def\mn@urlcharsother{\let\do\@makeother \do\$\do\&\do\#\do\^\do\_\do\%\do\~}
\def\mn@doi{\begingroup\mn@urlcharsother \@ifnextchar [ {\mn@doi@}
  {\mn@doi@[]}}
\def\mn@doi@[#1]#2{\def\@tempa{#1}\ifx\@tempa\@empty \href
  {http://dx.doi.org/#2} {doi:#2}\else \href {http://dx.doi.org/#2} {#1}\fi
  \endgroup}
\def\mn@eprint#1#2{\mn@eprint@#1:#2::\@nil}
\def\mn@eprint@arXiv#1{\href {http://arxiv.org/abs/#1} {{\tt arXiv:#1}}}
\def\mn@eprint@dblp#1{\href {http://dblp.uni-trier.de/rec/bibtex/#1.xml}
  {dblp:#1}}
\def\mn@eprint@#1:#2:#3:#4\@nil{\def\@tempa {#1}\def\@tempb {#2}\def\@tempc
  {#3}\ifx \@tempc \@empty \let \@tempc \@tempb \let \@tempb \@tempa \fi \ifx
  \@tempb \@empty \def\@tempb {arXiv}\fi \@ifundefined
  {mn@eprint@\@tempb}{\@tempb:\@tempc}{\expandafter \expandafter \csname
  mn@eprint@\@tempb\endcsname \expandafter{\@tempc}}}

\bibitem[\protect\citeauthoryear{{Abolfathi} et~al.,}{{Abolfathi}
  et~al.}{2018}]{Abolfathi2018}
{Abolfathi} B.,  et~al., 2018, \mn@doi [\apjs] {10.3847/1538-4365/aa9e8a},
  \href {http://adsabs.harvard.edu/abs/2018ApJS..235...42A} {235, 42}

\bibitem[\protect\citeauthoryear{{Alam} et~al.,}{{Alam}
  et~al.}{2015}]{Alam2015}
{Alam} S.,  et~al., 2015, \mn@doi [\apjs] {10.1088/0067-0049/219/1/12}, \href
  {http://cdsads.u-strasbg.fr/abs/2015ApJS..219...12A} {219, 12}

\bibitem[\protect\citeauthoryear{{Allen}, {Hewett}, {Maddox}, {Richards}  \&
  {Belokurov}}{{Allen} et~al.}{2011}]{Allen2011}
{Allen} J.~T.,  {Hewett} P.~C.,  {Maddox} N.,  {Richards} G.~T.,   {Belokurov}
  V.,  2011, \mn@doi [\mnras] {10.1111/j.1365-2966.2010.17489.x}, \href
  {http://cdsads.u-strasbg.fr/abs/2011MNRAS.410..860A} {410, 860}

\bibitem[\protect\citeauthoryear{{Capellupo}, {Hamann}, {Shields},
  {Rodr{\'{\i}}guez Hidalgo}  \& {Barlow}}{{Capellupo}
  et~al.}{2011}]{Capellupo2011}
{Capellupo} D.~M.,  {Hamann} F.,  {Shields} J.~C.,  {Rodr{\'{\i}}guez Hidalgo}
  P.,   {Barlow} T.~A.,  2011, \mn@doi [\mnras]
  {10.1111/j.1365-2966.2010.18185.x}, \href
  {http://cdsads.u-strasbg.fr/abs/2011MNRAS.413..908C} {413, 908}

\bibitem[\protect\citeauthoryear{{Chen}, {Pang}, {He}  \& {Huang}}{{Chen}
  et~al.}{2018a}]{Chen2018a}
{Chen} Z.-F.,  {Pang} T.-T.,  {He} B.,   {Huang} Y.,  2018a, \mn@doi [\apjs]
  {10.3847/1538-4365/aabcd4}, \href
  {http://cdsads.u-strasbg.fr/abs/2018ApJS..236...39C} {236, 39}

\bibitem[\protect\citeauthoryear{{Chen} et~al.,}{{Chen}
  et~al.}{2018b}]{Chen2018b}
{Chen} Z.-F.,  et~al., 2018b, \mn@doi [\apjs] {10.3847/1538-4365/aaeac3}, \href
  {http://cdsads.u-strasbg.fr/abs/2018ApJS..239...23C} {239, 23}

\bibitem[\protect\citeauthoryear{{Crenshaw}, {Kraemer}  \& {George}}{{Crenshaw}
  et~al.}{2003}]{Crenshaw2003}
{Crenshaw} D.~M.,  {Kraemer} S.~B.,   {George} I.~M.,  2003, \mn@doi [\araa]
  {10.1146/annurev.astro.41.082801.100328}, \href
  {http://cdsads.u-strasbg.fr/abs/2003ARA%26A..41..117C} {41, 117}

\bibitem[\protect\citeauthoryear{{Dawson} et~al.,}{{Dawson}
  et~al.}{2013}]{Dawson2013}
{Dawson} K.~S.,  et~al., 2013, \mn@doi [\aj] {10.1088/0004-6256/145/1/10},
  \href {http://cdsads.u-strasbg.fr/abs/2013AJ....145...10D} {145, 10}

\bibitem[\protect\citeauthoryear{{Eisenstein} et~al.,}{{Eisenstein}
  et~al.}{2011}]{Eisenstein2011}
{Eisenstein} D.~J.,  et~al., 2011, \mn@doi [\aj] {10.1088/0004-6256/142/3/72},
  \href {http://cdsads.u-strasbg.fr/abs/2011AJ....142...72E} {142, 72}

\bibitem[\protect\citeauthoryear{{Farrah}, {Lacy}, {Priddey}, {Borys}  \&
  {Afonso}}{{Farrah} et~al.}{2007}]{Farrah2007}
{Farrah} D.,  {Lacy} M.,  {Priddey} R.,  {Borys} C.,   {Afonso} J.,  2007,
  \mn@doi [\apjl] {10.1086/519492}, \href
  {http://cdsads.u-strasbg.fr/abs/2007ApJ...662L..59F} {662, L59}

\bibitem[\protect\citeauthoryear{{Ganguly} \& {Brotherton}}{{Ganguly} \&
  {Brotherton}}{2008}]{Ganguly2008}
{Ganguly} R.,  {Brotherton} M.~S.,  2008, \mn@doi [\apj] {10.1086/524106},
  \href {http://cdsads.u-strasbg.fr/abs/2008ApJ...672..102G} {672, 102}

\bibitem[\protect\citeauthoryear{{Gibson}, {Brandt}, {Schneider}  \&
  {Gallagher}}{{Gibson} et~al.}{2008}]{Gibson2008}
{Gibson} R.~R.,  {Brandt} W.~N.,  {Schneider} D.~P.,   {Gallagher} S.~C.,
  2008, \mn@doi [\apj] {10.1086/527462}, \href
  {http://cdsads.u-strasbg.fr/abs/2008ApJ...675..985G} {675, 985}

\bibitem[\protect\citeauthoryear{{Gibson} et~al.,}{{Gibson}
  et~al.}{2009}]{Gibson2009a}
{Gibson} R.~R.,  et~al., 2009, \mn@doi [\apj] {10.1088/0004-637X/692/1/758},
  \href {http://cdsads.u-strasbg.fr/abs/2009ApJ...692..758G} {692, 758}

\bibitem[\protect\citeauthoryear{{Grier} et~al.,}{{Grier}
  et~al.}{2015}]{Grier2015}
{Grier} C.~J.,  et~al., 2015, \mn@doi [\apj] {10.1088/0004-637X/806/1/111},
  \href {http://cdsads.u-strasbg.fr/abs/2015ApJ...806..111G} {806, 111}

\bibitem[\protect\citeauthoryear{{Hamann}, {Barlow}  \& {Junkkarinen}}{{Hamann}
  et~al.}{1997}]{Hamann1997}
{Hamann} F.,  {Barlow} T.~A.,   {Junkkarinen} V.,  1997, \mn@doi [\apj]
  {10.1086/303782}, \href {http://cdsads.u-strasbg.fr/abs/1997ApJ...478...87H}
  {478, 87}

\bibitem[\protect\citeauthoryear{{Hamann}, {Kaplan}, {Rodr{\'{\i}}guez
  Hidalgo}, {Prochaska}  \& {Herbert-Fort}}{{Hamann} et~al.}{2008}]{Hamann2008}
{Hamann} F.,  {Kaplan} K.~F.,  {Rodr{\'{\i}}guez Hidalgo} P.,  {Prochaska}
  J.~X.,   {Herbert-Fort} S.,  2008, \mn@doi [\mnras]
  {10.1111/j.1745-3933.2008.00554.x}, \href
  {http://cdsads.u-strasbg.fr/abs/2008MNRAS.391L..39H} {391, L39}

\bibitem[\protect\citeauthoryear{{He}, {Wang}, {Zhou}, {Bian}, {Liu}, {Yang},
  {Dou}  \& {Sun}}{{He} et~al.}{2017}]{He2017}
{He} Z.,  {Wang} T.,  {Zhou} H.,  {Bian} W.,  {Liu} G.,  {Yang} C.,  {Dou} L.,
   {Sun} L.,  2017, \mn@doi [\apjs] {10.3847/1538-4365/aa647a}, \href
  {http://cdsads.u-strasbg.fr/abs/2017ApJS..229...22H} {229, 22}

\bibitem[\protect\citeauthoryear{{Hemler} et~al.,}{{Hemler}
  et~al.}{2019}]{Hemler2019}
{Hemler} Z.~S.,  et~al., 2019, \mn@doi [\apj] {10.3847/1538-4357/aaf1bf}, \href
  {http://cdsads.u-strasbg.fr/abs/2019ApJ...872...21H} {872, 21}

\bibitem[\protect\citeauthoryear{{Hewett} \& {Wild}}{{Hewett} \&
  {Wild}}{2010}]{Hewett2010}
{Hewett} P.~C.,  {Wild} V.,  2010, \mn@doi [\mnras]
  {10.1111/j.1365-2966.2010.16648.x}, \href
  {http://cdsads.u-strasbg.fr/abs/2010MNRAS.405.2302H} {405, 2302}

\bibitem[\protect\citeauthoryear{{King}}{{King}}{2010}]{King2010}
{King} A.~R.,  2010, in {Maraschi} L.,  {Ghisellini} G.,  {Della Ceca} R.,
  {Tavecchio} F.,  eds,  Astronomical Society of the Pacific Conference Series
  Vol. 427, Accretion and Ejection in AGN: a Global View. p.~315

\bibitem[\protect\citeauthoryear{{Lu} \& {Lin}}{{Lu} \&
  {Lin}}{2018a}]{Lu2018complex1}
{Lu} W.-J.,  {Lin} Y.-R.,  2018a, \mn@doi [\mnras] {10.1093/mnras/stx2970},
  \href {http://cdsads.u-strasbg.fr/abs/2018MNRAS.474.3397L} {474, 3397}

\bibitem[\protect\citeauthoryear{{Lu} \& {Lin}}{{Lu} \&
  {Lin}}{2018b}]{Lu2018saturation}
{Lu} W.-J.,  {Lin} Y.-R.,  2018b, \mn@doi [\apj] {10.3847/1538-4357/aaca31},
  \href {http://cdsads.u-strasbg.fr/abs/2018ApJ...862...46L} {862, 46}

\bibitem[\protect\citeauthoryear{{Lu} \& {Lin}}{{Lu} \&
  {Lin}}{2018c}]{Lu2018complex2}
{Lu} W.-J.,  {Lin} Y.-R.,  2018c, \mn@doi [\apj] {10.3847/1538-4357/aad411},
  \href {http://cdsads.u-strasbg.fr/abs/2018ApJ...863..186L} {863, 186}

\bibitem[\protect\citeauthoryear{{Lu} et~al.,}{{Lu} et~al.}{2017}]{Lu2017}
{Lu} W.-J.,  et~al., 2017, \mn@doi [\mnras] {10.1093/mnrasl/slx013}, \href
  {http://cdsads.u-strasbg.fr/abs/2017MNRAS.468L...6L} {468, L6}

\bibitem[\protect\citeauthoryear{Lu, Lin  \& Qin}{Lu
  et~al.}{2018}]{Lu2018broad}
Lu W.-J.,  Lin Y.-R.,   Qin Y.-P.,  2018, \mn@doi [\mnras]
  {10.1093/mnrasl/slx176}, 473, L106

\bibitem[\protect\citeauthoryear{{Misawa}, {Charlton}, {Eracleous}, {Ganguly},
  {Tytler}, {Kirkman}, {Suzuki}  \& {Lubin}}{{Misawa}
  et~al.}{2007a}]{Misawa2007a}
{Misawa} T.,  {Charlton} J.~C.,  {Eracleous} M.,  {Ganguly} R.,  {Tytler} D.,
  {Kirkman} D.,  {Suzuki} N.,   {Lubin} D.,  2007a, \mn@doi [\apjs]
  {10.1086/513713}, \href {http://cdsads.u-strasbg.fr/abs/2007ApJS..171....1M}
  {171, 1}

\bibitem[\protect\citeauthoryear{{Misawa}, {Eracleous}, {Charlton}  \&
  {Kashikawa}}{{Misawa} et~al.}{2007b}]{Misawa2007b}
{Misawa} T.,  {Eracleous} M.,  {Charlton} J.~C.,   {Kashikawa} N.,  2007b,
  \mn@doi [\apj] {10.1086/513097}, \href
  {http://cdsads.u-strasbg.fr/abs/2007ApJ...660..152M} {660, 152}

\bibitem[\protect\citeauthoryear{{Misawa}, {Charlton}  \& {Eracleous}}{{Misawa}
  et~al.}{2014}]{Misawa2014}
{Misawa} T.,  {Charlton} J.~C.,   {Eracleous} M.,  2014, \mn@doi [\apj]
  {10.1088/0004-637X/792/1/77}, \href
  {http://cdsads.u-strasbg.fr/abs/2014ApJ...792...77M} {792, 77}

\bibitem[\protect\citeauthoryear{{Schneider} et~al.,}{{Schneider}
  et~al.}{1993}]{Schneider1993}
{Schneider} D.~P.,  et~al., 1993, \mn@doi [\apjs] {10.1086/191798}, \href
  {http://cdsads.u-strasbg.fr/abs/1993ApJS...87...45S} {87, 45}

\bibitem[\protect\citeauthoryear{{Shen} et~al.,}{{Shen}
  et~al.}{2015}]{Shen2015}
{Shen} Y.,  et~al., 2015, VizieR Online Data Catalog, \href
  {http://cdsads.u-strasbg.fr/abs/2015yCat..22160004S} {221}

\bibitem[\protect\citeauthoryear{{Shi}, {Jiang}, {Wang}, {Zhang}, {Ji}, {Liu}
  \& {Zhou}}{{Shi} et~al.}{2016}]{Shi2016}
{Shi} X.-H.,  {Jiang} P.,  {Wang} H.-Y.,  {Zhang} S.-H.,  {Ji} T.,  {Liu}
  W.-J.,   {Zhou} H.-Y.,  2016, \mn@doi [\apj] {10.3847/0004-637X/829/2/96},
  \href {http://cdsads.u-strasbg.fr/abs/2016ApJ...829...96S} {829, 96}

\bibitem[\protect\citeauthoryear{{Smee} et~al.,}{{Smee}
  et~al.}{2013}]{Smee2013}
{Smee} S.~A.,  et~al., 2013, \mn@doi [\aj] {10.1088/0004-6256/146/2/32}, \href
  {http://cdsads.u-strasbg.fr/abs/2013AJ....146...32S} {146, 32}

\bibitem[\protect\citeauthoryear{{Springel}, {Di Matteo}  \&
  {Hernquist}}{{Springel} et~al.}{2005}]{Springel2005}
{Springel} V.,  {Di Matteo} T.,   {Hernquist} L.,  2005, \mn@doi [\apjl]
  {10.1086/428772}, \href {http://cdsads.u-strasbg.fr/abs/2005ApJ...620L..79S}
  {620, L79}

\bibitem[\protect\citeauthoryear{{Vivek}, {Srianand}, {Petitjean}, {Mohan},
  {Mahabal}  \& {Samui}}{{Vivek} et~al.}{2014}]{Vivek2014}
{Vivek} M.,  {Srianand} R.,  {Petitjean} P.,  {Mohan} V.,  {Mahabal} A.,
  {Samui} S.,  2014, \mn@doi [\mnras] {10.1093/mnras/stu288}, \href
  {http://cdsads.u-strasbg.fr/abs/2014MNRAS.440..799V} {440, 799}

\bibitem[\protect\citeauthoryear{{Wang}, {Yang}, {Wang}  \& {Ferland}}{{Wang}
  et~al.}{2015}]{Wang2015}
{Wang} T.,  {Yang} C.,  {Wang} H.,   {Ferland} G.,  2015, \mn@doi [\apj]
  {10.1088/0004-637X/814/2/150}, \href
  {http://cdsads.u-strasbg.fr/abs/2015ApJ...814..150W} {814, 150}

\bibitem[\protect\citeauthoryear{{Weymann}, {Morris}, {Foltz}  \&
  {Hewett}}{{Weymann} et~al.}{1991}]{Weymann1991}
{Weymann} R.~J.,  {Morris} S.~L.,  {Foltz} C.~B.,   {Hewett} P.~C.,  1991,
  \mn@doi [\apj] {10.1086/170020}, \href
  {http://cdsads.u-strasbg.fr/abs/1991ApJ...373...23W} {373, 23}

\bibitem[\protect\citeauthoryear{{Wildy}, {Goad}  \& {Allen}}{{Wildy}
  et~al.}{2014}]{Wildy2014}
{Wildy} C.,  {Goad} M.~R.,   {Allen} J.~T.,  2014, \mn@doi [\mnras]
  {10.1093/mnras/stt2028}, \href
  {http://cdsads.u-strasbg.fr/abs/2014MNRAS.437.1976W} {437, 1976}

\makeatother
\end{thebibliography}


\clearpage
\appendix
\begin{table}
   \centering
   \caption{The fitting data of BAL and the power-law continuum spectra.}
	\label{tab:1}
	\begin{tabular}{cccccc} 
		\hline
           MJD   &  S/N &  $EW_{BAL-L}  $   &  $EW_{BAL-R}  $  & $EW_{BAL}  $  &  $F $ \\
         
 (Day)&    & $(${\AA}$)$&   $(${\AA}$)$&	 $(${\AA}$)$&	 ($10^{-13} erg cm^{-2} s^{-1}$)\\
 \hline
56426& 23.87 & 2.67$\pm$0.29&  1.45$\pm$0.22&	4.13$\pm$0.36&	1.7184$\pm$0.0017\\
56660& 27.32 & 2.96$\pm$0.28&	1.59$\pm$0.07&	4.54$\pm$0.29&	1.5084$\pm$0.0013\\
56664& 23.39 & 2.79$\pm$0.37&	1.54$\pm$0.08&	4.33$\pm$0.38&	1.5297$\pm$0.0015\\
56683& 29.40 & 2.60$\pm$0.25&	1.61$\pm$0.06&	4.20$\pm$0.26&	1.7184$\pm$0.0010\\
56686& 29.05 & 2.72$\pm$0.22&	1.59$\pm$0.06&	4.31$\pm$0.23&	1.4409$\pm$0.0017\\
56697& 28.84 & 2.94$\pm$0.06&	1.98$\pm$0.06&	4.92$\pm$0.08&	1.3859$\pm$0.0011\\
56717& 32.71 & 3.12$\pm$0.21&	2.01$\pm$0.06&	5.12$\pm$0.22&	1.4427$\pm$0.0010\\
56720& 32.05 & 3.32$\pm$0.23&	2.26$\pm$0.06&	5.57$\pm$0.24&	1.4859$\pm$0.0011\\
56722& 37.00 & 3.33$\pm$0.06&	1.85$\pm$0.05&	5.18$\pm$0.07&	1.5376$\pm$0.0010\\
56726& 31.22 & 3.01$\pm$0.06&	1.86$\pm$0.06&	5.56$\pm$0.08&	1.3952$\pm$0.0011\\
56739& 29.77 & 2.84$\pm$0.07&	1.83$\pm$0.07&	4.67$\pm$0.10&	1.4124$\pm$0.0011\\
56745& 29.96 & 2.62$\pm$0.07&	1.55$\pm$0.06&	4.17$\pm$0.09&	1.5610$\pm$0.0013\\
56749& 29.96 & 2.88$\pm$0.06&	1.70$\pm$0.21&	4.17$\pm$0.22&	1.3368$\pm$0.0010\\
56751& 31.51 & 2.86$\pm$0.22&	1.93$\pm$0.06&	4.78$\pm$0.23&	1.4124$\pm$0.0011\\
56755& 29.75 & 2.20$\pm$0.17&	1.51$\pm$0.07&	3.70$\pm$0.18&	1.4442$\pm$0.0011\\
56768& 32.17 & 2.72$\pm$0.06&	1.62$\pm$0.16&	4.34$\pm$0.17&	1.3348$\pm$0.0009\\
56772& 37.54 & 2.29$\pm$0.16&	1.44$\pm$0.05&	3.86$\pm$0.16&	1.6430$\pm$0.0010\\
56780& 30.48 & 2.28$\pm$0.21&	1.55$\pm$0.06&	3.83$\pm$0.22&	1.4145$\pm$0.0010\\
56782& 36.71 & 2.18$\pm$0.13&	1.68$\pm$0.05&	3.86$\pm$0.14&	1.5922$\pm$0.0011\\
56795& 30.16 & 2.23$\pm$0.23&	1.47$\pm$0.07&	3.69$\pm$0.24&	1.5701$\pm$0.0012\\
56799& 37.04 & 2.21$\pm$0.16&	1.37$\pm$0.06&	3.57$\pm$0.17&	1.5560$\pm$0.0010\\
56804& 29.50 & 1.27$\pm$0.23&	0.66$\pm$0.13&	1.94$\pm$0.26&	1.6163$\pm$0.0013\\
56808& 30.47 & 1.79$\pm$0.18&	1.36$\pm$0.07&	3.15$\pm$0.19&	1.5575$\pm$0.0012\\
56825& 33.75 & 2.15$\pm$0.24&	1.17$\pm$0.19&	3.31$\pm$0.30&	1.5660$\pm$0.0011\\
56829& 31.70 & 1.84$\pm$0.23&	0.93$\pm$0.14&	2.77$\pm$0.27&	1.5145$\pm$0.0012\\
56833& 31.71 & 1.80$\pm$0.18&	0.78$\pm$0.16&	2.57$\pm$0.24&	1.4876$\pm$0.0012\\
56837& 35.22 & 1.42$\pm$0.14&	1.01$\pm$0.16&	2.43$\pm$0.21&	1.4932$\pm$0.0011\\
57038& 25.30 & 1.66$\pm$0.16&	0.95$\pm$0.14&	2.60$\pm$0.21&	1.4941$\pm$0.0013\\
57050& 25.08 & 1.45$\pm$0.27&	0.86$\pm$0.23&	2.32$\pm$0.35&	1.5496$\pm$0.0014\\
57067& 22.59 & 1.43$\pm$0.08&	1.01$\pm$0.09&	2.44$\pm$0.12&	1.4071$\pm$0.0014\\
57082& 22.94 & 1.32$\pm$0.19&	1.04$\pm$0.32&	2.36$\pm$0.37&	1.3067$\pm$0.0013\\
57106& 24.77 & 1.34$\pm$0.08&	0.93$\pm$0.09&	2.29$\pm$0.16&	1.1529$\pm$0.0012\\
57127& 25.03 & 1.27$\pm$0.25&	0.95$\pm$0.34&	2.22$\pm$0.42&	1.4575$\pm$0.0014\\
57135& 25.95 & 1.51$\pm$0.35&	0.96$\pm$0.07&	2.47$\pm$0.35&	1.4026$\pm$0.0012\\
57159& 24.77 & 1.59$\pm$0.32&	0.89$\pm$0.19&	2.48$\pm$0.38&	1.3938$\pm$0.0013\\
57166& 26.58 & 1.59$\pm$0.35&	1.18$\pm$0.25&	2.77$\pm$0.43&	1.4014$\pm$0.0012\\
57185& 24.40 & 0.96$\pm$0.14&	0.38$\pm$0.08&	1.34$\pm$0.16&	1.4422$\pm$0.0014\\
57196& 26.86 & 1.30$\pm$0.07&	0.64$\pm$0.11&	1.94$\pm$0.13&	1.4273$\pm$0.0012\\
57428& 22.23 & 1.66$\pm$0.10&	1.06$\pm$0.28&	2.72$\pm$0.3&	1.4876$\pm$0.0015\\
57435& 25.29 & 0.99$\pm$0.15&	0.49$\pm$0.08&	1.50$\pm$0.18&	1.4542$\pm$0.0013\\
57451& 25.66 & 1.35$\pm$0.08&	1.08$\pm$0.21&	2.43$\pm$0.22&	1.4725$\pm$0.0013\\
57463& 23.50 & 1.45$\pm$0.26&	1.31$\pm$0.08&	2.76$\pm$0.27&	1.4697$\pm$0.0014\\
57492& 22.80 & 1.56$\pm$0.10&	0.75$\pm$0.09&	2.32$\pm$0.14&	1.4700$\pm$0.0014\\
57510& 26.29 & 1.32$\pm$0.17&	1.24$\pm$0.08&	2.56$\pm$0.19&	1.4565$\pm$0.0014\\
		\hline
	\end{tabular}
  \footnotesize
\end{table}

\begin{table*}
   \centering
\caption{Correlated parameters of EW versus F and EW versus EW for each sample.}
	\label{tab:2}
	\begin{tabular}{ccccccccc} 
		\hline
        \hline
		    &    EW VS. F   & &  &      EW VS. F &  &  &     EW VS. F &  \\
            &  for MJD 56426-56837 & &    &for MJD 57038-57510 & &     & for All sample  &\\
		\hline
            &  r$^a$ & p$^b$       &   & r & p                  &  & r  & p\\		
		BAL    &   -0.46  &	0.016  &   & 0.12    & 	0.653      &  &  0.06    &	0.718 \\
		BAL-L & -0.47  & 	0.014  &   &	0.25   &  	0.333      &  & 0.08	&0.622	    \\
        BAL-R & -0.54  &	0.003  &   &	0.03    &	0.896      &  & -0.01    & 	0.952	\\
        \hline
        
        & EW VS. EW &   &  & \\ 
        \hline
        & r & p &   &  \\
	L-R & 0.88 & $4.044\times 10^{-15}$ &   &  \\
		\hline
		\hline
	\end{tabular}
    
    \begin{tablenotes}
  \footnotesize
  \item \emph{Notes.}$^{\rm a}$The Spearman rank correlation coefficient.
  \item $^{\rm b}$The significance level of the correlation coefficient.
  \end{tablenotes}
\end{table*}


\bsp    
\label{lastpage}
\end{document}